\documentclass{iau}

\usepackage{amsmath}
\usepackage{graphicx}
\usepackage{multirow}

\begin{document}

\lefttitle{Cambridge Author}
\righttitle{Septuple Cluster System and Extended Family}

\jnlPage{1}{7}
\jnlDoiYr{2025}
\doival{10.1017/xxxx}

\aopheadtitle{Proceedings IAU Symposium}
\editors{H. M. Lee, R. Spurzem \& J. Hong, eds.}

\title{Analysis of a Septuple Open Cluster System and Its Extended Family in Gaia DR3}

\author{Muhammad Akmal Husain$^{1}$, Ferdinand$^2$, Mochamad Ikbal Arifyanto$^{3,4}$, Muhammad Irfan Hakim$^{3,4}$}

\affiliation{
	$^1$ Computational Science Graduate Program, Faculty of Mathematics and Natural Sciences, Bandung Institute of Technology, Bandung 40132, Indonesia \\
	$^2$ Department of Astronomy, University of Illinois Urbana-Champaign, IL 61820, USA \\
	$^3$ Astronomy Research Group, Faculty of Mathematics and Natural Sciences, Bandung Institute of Technology, Bandung 40132, Indonesia \\
	$^4$ Bosscha Observatory, Bandung Institute of Technology, Jl. Peneropongan Bintang No. 1, West Java 40391, Indonesia
}

\begin{abstract}
A rare multiple open cluster system has been analyzed using Gaia DR3 astrometry and photometry data. Using Agglomerative Hierarchical clustering and Bayesian-HDBSCAN*, we identify a compact core consisting of seven known open clusters and two additional components, including a new candidate, forming a nine-member association. Membership probabilities are refined through statistical modeling, combining GUMM, Bayesian-KDE, and Bayesian-XDGMM for tidal tails identification. Backward orbital integrations confirm coherent motions over 20–30 million years, suggesting a common origin from the same massive molecular cloud. This system offers a unique laboratory for investigating cluster multiplicity, dynamical evolution, and Galactic structure.
\end{abstract}


\maketitle

\section{Introduction}

Open clusters can provide insights into the structure and evolution of galaxies. Valuable information about star formation, stellar evolution, and the chemical enrichment of the interstellar medium can be obtained from these objects, which typically form from a single molecular cloud \citep{Soubiran2018}. Their similar composition and age make them excellent reference objects for testing theoretical models of stellar evolution and internal dynamics \citep{Randich2018}. High-precision astrometric measurements from Gaia can identify cluster membership, reconstruct spatial distributions, evaluate interior kinematics, and reliably measure ages \citep{Hunt2023}.

Double and multiple open cluster systems are invaluable as laboratories for studying environmental effects and gravitational interactions between clusters. These systems provide insights into whether these clusters originated from a single massive molecular cloud \citep{Song2022, Qin2022, Palma2025}. Gaia Data Release 3 (hereafter Gaia DR3) has identified more pairs of connected clusters, demonstrating the complex processes of cluster formation and evolution in the Galaxy \citep{Almeida2023, Zhong2022}. Furthermore, these systems serve as references for refining models of the potential, mass distribution, and chemical enrichment in our Galaxy \citep{Ferreira2020}. This study focuses on the identification of the membership, structural evolution, and dynamics of several open cluster systems within the Milky Way, utilizing Gaia DR3 data to investigate the impact of gravitational interactions on their evolution and their contribution to the Galaxy's stellar population.

\section{Data and Methods}

To search for double or multiple open clusters, we applied the Agglomerative Hierarchical clustering method to the catalog \citet{Hunt2023} using three-dimensional Galactocentric coordinates with a separation threshold of 50 pc. The results obtained are used as a reference point for the clustering pipeline we have designed to ensure that the clustering is performed on this cluster system. We designed clustering pipeline on Gaia DR3 astrometric data (RA, Dec, $\mu_{\alpha^{*}}$, $\mu_{\delta}$, $\varpi$) using HDBSCAN* \citep{McInnes2017} with a Bayesian approach. Given the proximity of clusters, which can cause cross-contamination where stars may be assigned to multiple clusters simultaneously, the Bayesian approach is crucial for accurate membership assignment. After initial clustering, the field is divided into three subregions to maximize the capabilities of HDBSCAN*. For each region, 1000 Monte Carlo iterations were performed, perturbed with astrometric vectors within the measurement uncertainty bounds, with Bayesian-HDBSCAN* applied using optimized parameters (\texttt{min\_cluster\_size} = 3, \texttt{min\_samples} = 4). Stars appearing in the same group in $\geq$50\% are classified as members of a single cluster. To minimize background star contamination, we apply the Gaussian-Uniform Mixture Model (GUMM) to model the cluster core as a multivariate Gaussian distribution against a uniform background, calculating posterior membership probabilities via maximum likelihood estimation. Adaptive thresholding using statistical criteria (elbow method, Ripley's $K$ function \citep{Ripley1976}) ensures spatial and probabilistic cohesion, while Bayesian Kernel Density Estimation (Bayesian-KDE) further refines membership by evaluating local density contrasts in astrometric parameter space. Extended stellar features, particularly tidal tails, are identified using eXtreme Deconvolution Gaussian Mixture Modeling \citep{Holoien2017} with a Bayesian approach (Bayesian-XDGMM) that combines astrometric and photometric likelihoods. The Bayesian framework is crucial here because overlapping cluster boundaries can cause individual stars to be incorrectly classified into multiple clusters, with stars exceeding the 75th percentile in posterior probability retained as high-confidence members.

Orbital reconstruction used the \texttt{gala} dynamics package \citep{PriceWhelan2017} with the \texttt{MilkyWayPotential2022} model. For each cluster, 1000 Monte Carlo samples of six-dimensional phase-space vectors (including Gaia astrometry and radial velocities) were integrated backward using symplectic integrators according to stellar ages from isochrone fitting with the \texttt{AsTeCA} code \citep{Perren2015}. This provided probabilistic orbital histories to assess the clusters' collective motion and long-term coherence, robustly evaluating physical connections within this multi-member system.

\section{Result and Discussions}

Our cluster analysis reveals a rare nine-member open cluster system, consisting of ASCC 19, ASCC 20, ASCC 21, Briceño 1, OC 0339, OCSN 61, UBC 17a, UBC 17b/Theia 13, and the newly identified UBC Family 17. Initially, Agglomerative Clustering in 3D Cartesian space using the \citet{Hunt2023} catalog identified seven clusters forming a compact core subsystem with distances of 13 to 47 parsecs, bounded within a projection radius of 5$^\circ$ centered at RA = 83$^\circ$, Dec = +1$^\circ$. Further processing revealed two additional clusters, UBC 17b/Theia 13 and UBC 17 Family, located slightly further from the center of the seven-cluster system but still closely connected to the OCSN 61 cluster, with an average distance of only about 21 parsecs.

\begin{figure}[!htb]
	\centering
	\begin{subfigure}{0.32\textwidth}
		\centering
		\includegraphics[width=\textwidth]{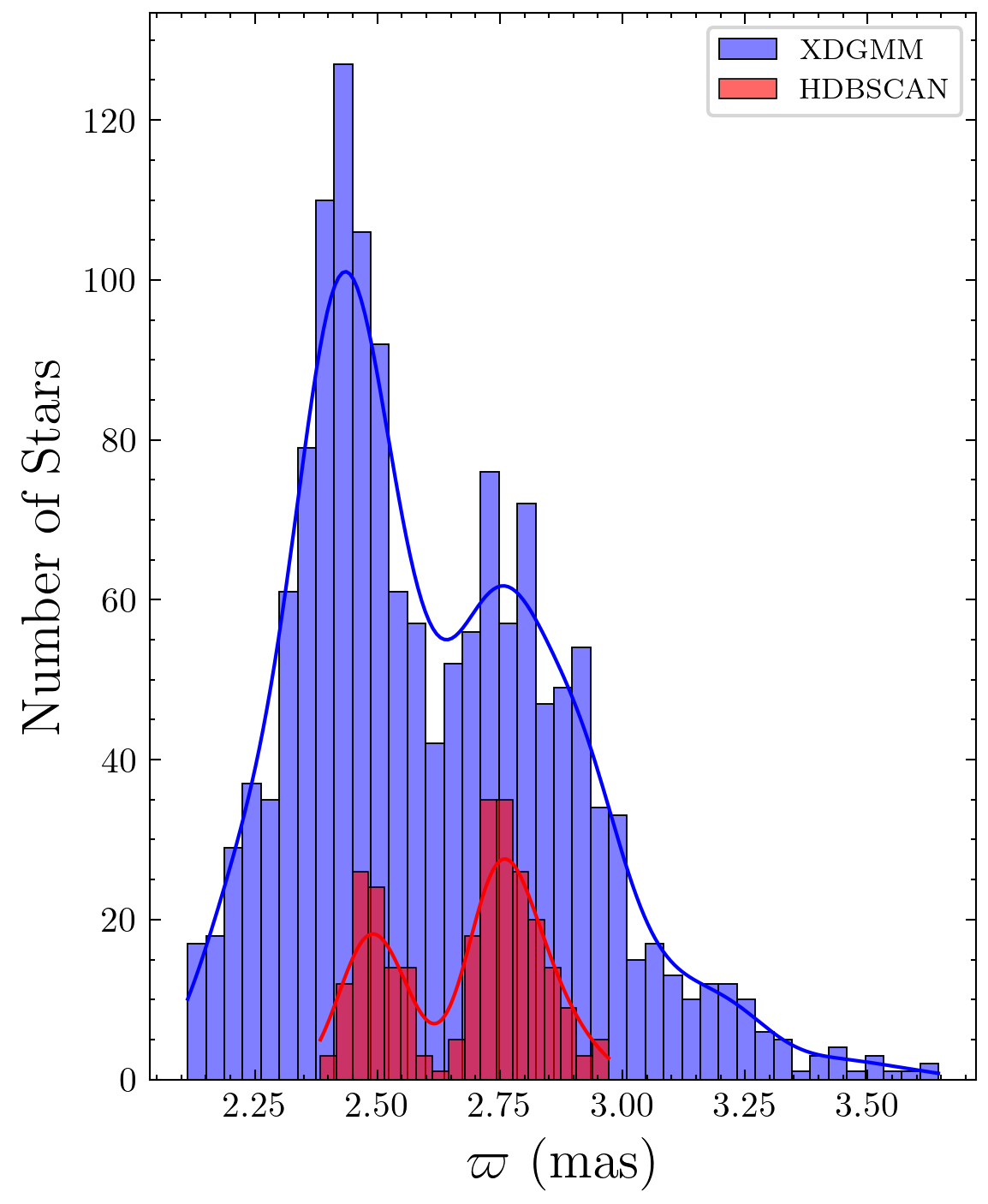}
		\caption{Bimodal parallax in UBC 17a.}
		\label{fig:Dist_parallax}
	\end{subfigure}
	\hfill
	\begin{subfigure}{0.32\textwidth}
		\centering
		\includegraphics[width=\textwidth]{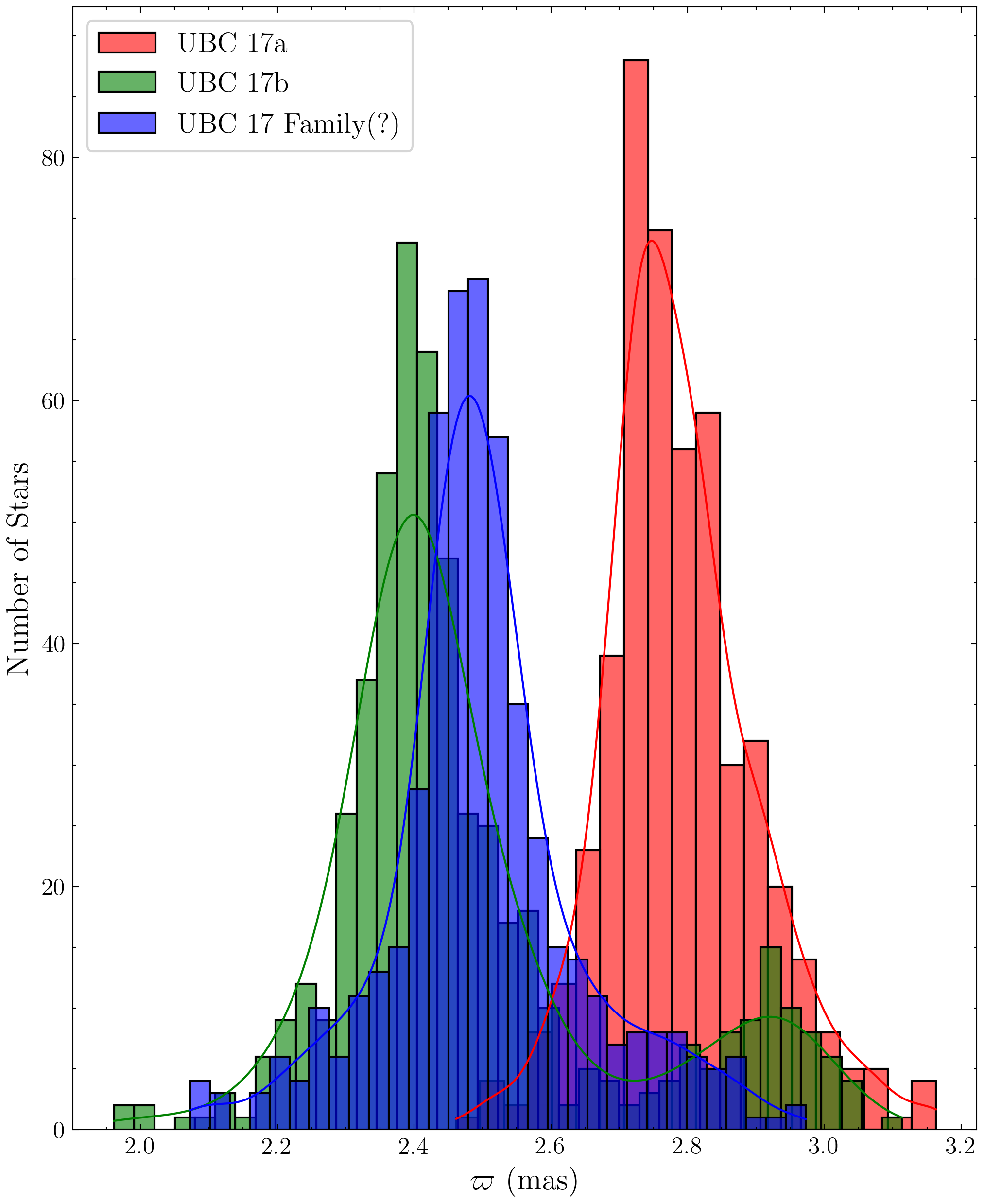}
		\caption{UBC 17a separation via GMM.}
		\label{fig:PM_diagram}
	\end{subfigure}
	\hfill
	\begin{subfigure}{0.32\textwidth}
		\centering
		\includegraphics[width=\textwidth]{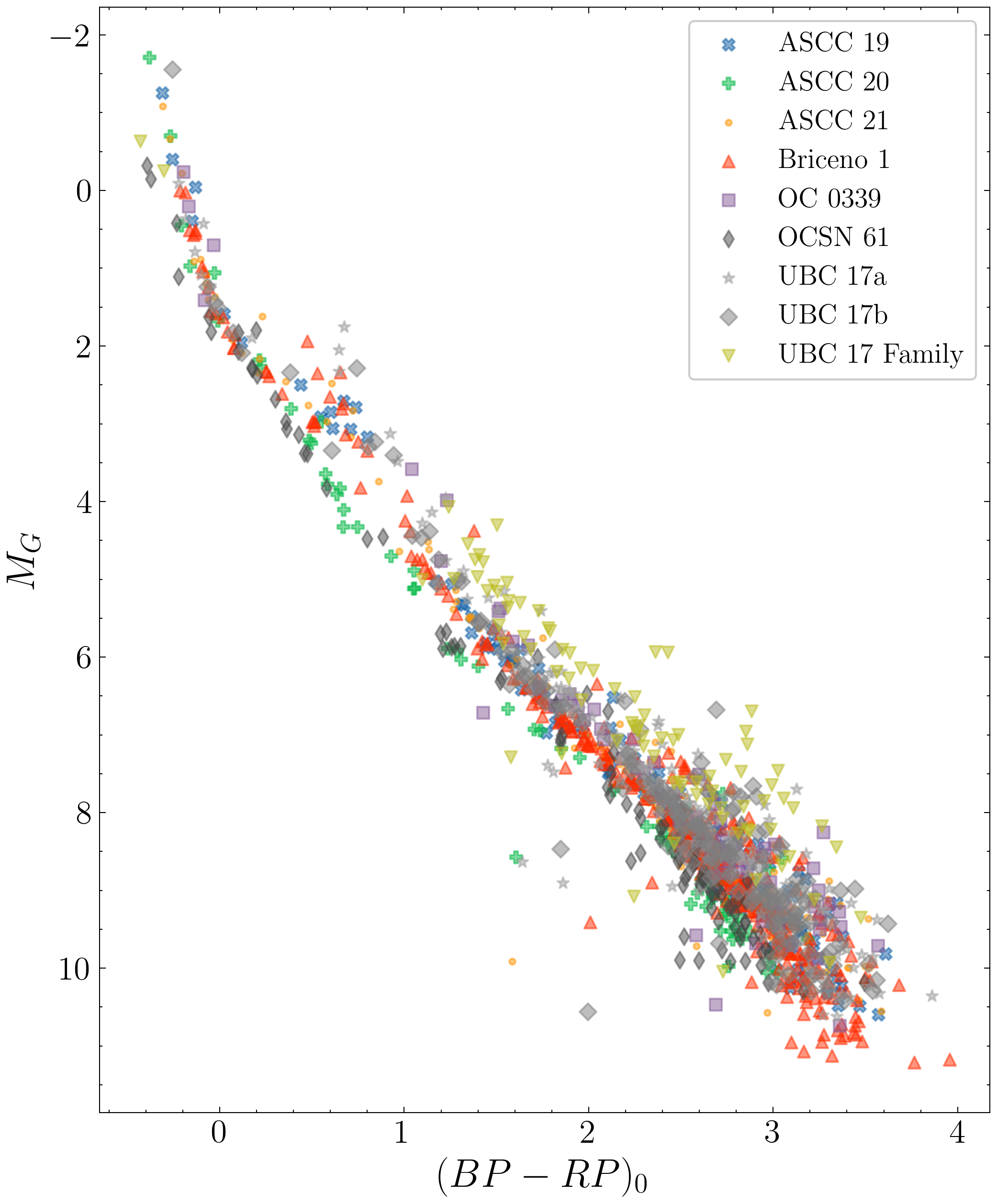}
		\caption{CMD showing coeval members.}
		\label{fig:CMD}
	\end{subfigure}
	
	\vspace{0.5cm}
	
	\begin{subfigure}{0.45\textwidth}
		\centering
		\includegraphics[width=\textwidth]{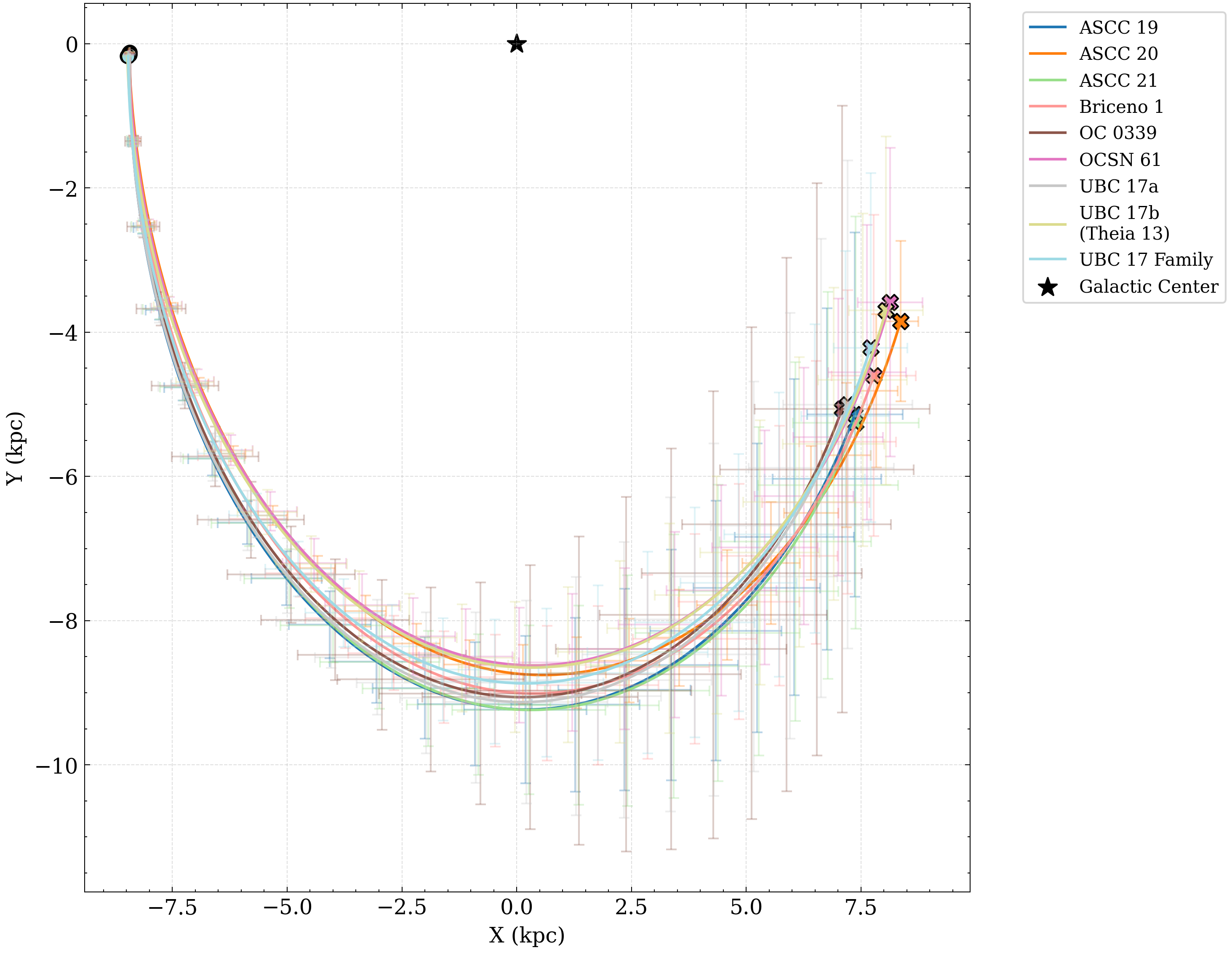}
		\caption{XY orbits with uncertainty.}
		\label{fig:fourth}
	\end{subfigure}
	\begin{subfigure}{0.4\textwidth}
		\centering
		\includegraphics[width=\textwidth]{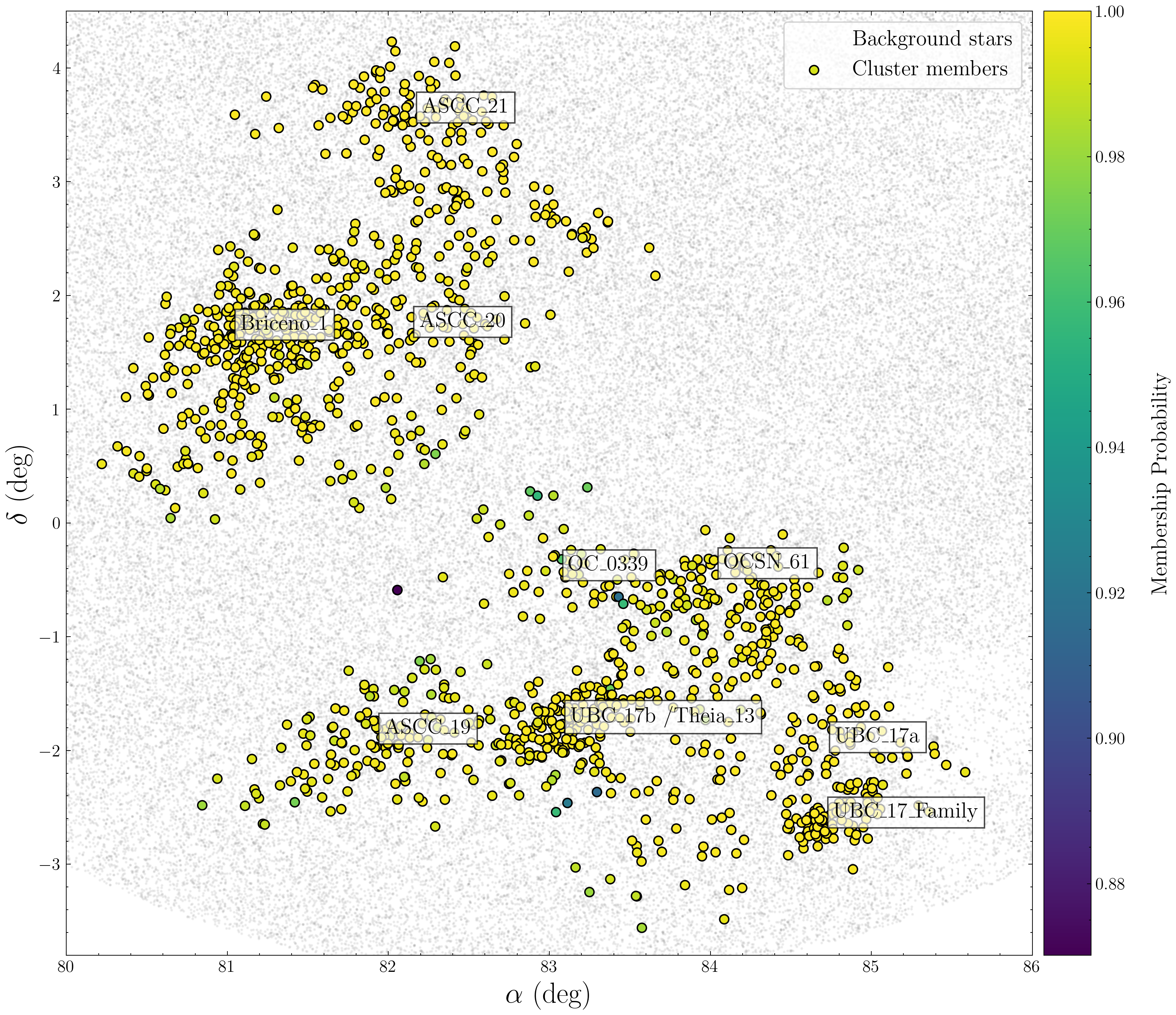}
		\caption{RA-Dec with membership scores.}
		\label{fig:fifth}
	\end{subfigure}
	
	\caption{Summary clustering analysis: (a) parallax distributionn of UBC 17a, (b) separating UBC 17a via GMM, (c) CMD of nine clusters, (d) orbital motion of nine clusters, (e) membership probability of nine clusters.}
	\label{fig:combined}
\end{figure}

Although the distance between these two additional clusters is closer to OCSN 61, their astrometric characteristics are more similar to UBC 17a, which is identified as a tidal tail (XDGMM result) as shown in the bimodal parallax distribution in Figure~\ref{fig:Dist_parallax}. After separation by the Gaussian Mixture Model (GMM) this results in three cluster components, namely UBC 17a, UBC 17b/Theia 13, and UBC 17 Family, shown in Figure~\ref{fig:PM_diagram}. This shows that these three components still have dense and bound cores, while their outer parts have started to expand and form filaments, indicating that they are in the decay stage. The spatial distribution of these nine clusters, shown in Fig.~\ref{fig:fifth}, shows a dense RA-Dec clustering with a high membership probability (average = 0.95), complemented by a strong kinematic alignment across the components, indicating a history of shared dynamics despite the separation. This illustrates how complex Galactic dynamics can extend into this system, in agreement with N-body simulations \citep{Gieles2011, Dalessandro2015}.

High-multiplicity binary open cluster systems are relatively rare in the Milky Way. Although recent Gaia-based studies have cataloged hundreds of binary clusters \citep{Piecka2021,Palma2025}.
\citet{Qin2023} found only 45 binaries but only three triplets within 500 pc of the Sun indicating that higher order systems such as triples and quadruples are very rare. Moreover, systems exceeding six members are increasingly rare, with examples including the seven clusters moving together in LISCA I \citep{Dalessandro2021} and the eight-member system reported by \citet{Palma2025}. Our discovery of nine members thus ranks among the highest-multiplicity associations known, highlighting their value for studying hierarchical star formation and the dynamical evolution of the Galaxy.

Photometric analysis further supports a common origin, with all nine clusters aligning on identical sequences of color-magnitude diagrams (hereafter CMD) (Figure~\ref{fig:CMD}) corrected using \citet{Casagrande2018} reddening calibration. Isochrone fitting yield ages of 16 to 32 million years ($\log(\rm age/year) = 7.20 \pm 0.15$) and nearly uniform metallicities ([Fe/H] = $0.018-0.060$). The coevolutionary loci of this system suggests they are formed from the same massive molecular cloud that maintained aggregation opportunities despite spatial separation. This suggests birth in hierarchically structured filaments, followed by outgassing and tidal release consistent with the turbulent fragmentation model \citep{Elmegreen1996, Krumholz2019}.

The orbital integration shown in Figure~\ref{fig:fourth} provides dynamical confirmation of this nine-cluster system. The backward integration reveals a common trajectory 20 to 30 million years ago, with an overlap of 1$\sigma$ including the distant components. The hierarchical configuration suggests the development of a compact core (seven clusters at an average distance of 25 parsecs) that maintains the original gravitational potential. OCSN 61 serves as a connecting population with the distant tidal pairs UBC 17b/Theia 13 and the UBC 17 Family. This structure exhibits progressive tidal stretching by Galactic forces over 20 million years while maintaining kinematic and chemical coherence, providing compelling empirical validation for cluster disruption models in the galactic potential.

\section{Acknowledgement }
MAH gratefully acknowledges the Leids Kerkhoven Bosscha Fonds (LKBF) for the travel grant supporting attendance at IAU Symposium 398. This work is also supported by the 2024 PPMI KK Astronomy Program, FMIPA, ITB.

\end{document}